\documentclass{ws-procs975x65}

\usepackage{graphicx}

\def\la{\mathrel{\mathchoice {\vcenter{\offinterlineskip\halign{\hfil
$\displaystyle##$\hfil\cr<\cr\sim\cr}}}
{\vcenter{\offinterlineskip\halign{\hfil$\textstyle##$\hfil\cr
<\cr\sim\cr}}}
{\vcenter{\offinterlineskip\halign{\hfil$\scriptstyle##$\hfil\cr
<\cr\sim\cr}}}
{\vcenter{\offinterlineskip\halign{\hfil$\scriptscriptstyle##$\hfil\cr
<\cr\sim\cr}}}}}

\begin{document}

\title{Magnetic  White   Dwarfs:  Observations,  Theory,   and  Future
  Prospects}

\author{Enrique Garc\'\i a--Berro$^{1,2,*}$} 

\address{$^1$Departament de F\'\i sica, 
             Universitat Polit\`ecnica de Catalunya,
             c/Esteve Terrades 5,
             08860 Castelldefels, Spain\\
         $^2$Institute for Space Studies of Catalonia, 
             c/Gran Capit\`a 2--4, 
             Edif. Nexus 201, 
             08034 Barcelona, Spain\\
         $^*$E-mail: enrique.garcia-berro@upc.edu}

\author{Mukremin Kilic}

\address{Homer L. Dodge Department of Physics and Astronomy,
         440 W. Brooks Street, 
         Norman, OK 73019,
         USA}

\author{S. O. Kepler}

\address{Instituto de Fisica, 
         Universidade Federal do Rio Grande do Sul, 
         91501-970 Porto-Alegre, RS, 
         Brazil}

\begin{abstract}
Isolated  magnetic  white dwarfs  have  field  strengths ranging  from
$10^3$~G  to   $10^9$~G,  and  constitute  an   interesting  class  of
objects. The  origin of the magnetic  field is still the  subject of a
hot debate.   Whether these fields  are fossil, hence the  remnants of
original  weak magnetic  fields  amplified during  the  course of  the
evolution of the  progenitor of white dwarfs, or on  the contrary, are
the  result  of  binary   interactions  or,  finally,  other  physical
mechanisms that  could produce such  large magnetic fields  during the
evolution of  the white  dwarf itself, remains  to be  elucidated.  In
this  work we  review the  current  status and  paradigms of  magnetic
fields in  white dwarfs, from  both the theoretical  and observational
points of view.
\end{abstract}


\bodymatter

\section{Introduction}	

Isolated  magnetic  white dwarfs  have  field  strengths ranging  from
$10^3$ to  $10^9$~G, and  are about  10\% of  the total  population of
single  white dwarfs,  although the  precise percentage  is still  the
subject  of  some  debate.   Specifically, the  percentage  of  single
magnetic  white  dwarfs   in  volume-limited  surveys\cite{Sion14}  is
typically 15\%,  whereas in  magnitude-limited samples\cite{Liebert03}
this percentage  decreases to about  4\%.  The number of  white dwarfs
with well determined magnetic fields has increased noticeably with the
advent of large  scale surveys, of which the Sloan  Digital Sky Survey
(SDSS)\cite{SDSS} is the leading example. The pioneering detections of
magnetic fields in  single white dwarfs\cite{Kuiper} were  done in the
mid  thirties,  and opened  a  new  field  of research.   These  early
discoveries  were  followed  by  more studies,  which  allowed  us  to
increase  the sample  of single  white dwarfs  with measured  magnetic
fields to about a few dozens.  However, from an observational point of
view, the largest breakthrough in the search for magnetic white dwarfs
arrived  with the  advent  large, automatic,  systematic surveys.   In
particular, the  SDSS has allowed us  to unveil a population  of about
600 magnetic white dwarfs\cite{Kepler15}.

Despite being  this an interesting  field of research, because  of its
many applications to other research areas  --- of which we mention the
field  of cataclysmic  variables, to  give  just one  example ---  the
impressive advance in the observational side, has not been followed by
theory, which remains  one step behind.  This lag is  partially due to
the intrinsic difficulty of modeling magnetic fields. Indeed, modeling
magnetic fields  is a tough  endeavour, as  in most cases  it requires
full three-dimensional simulations. Consequently, in many applications
crude simplifications are  done. However, this is not  the only reason
why  we still  do not  have a  comprehensive and  complete picture  of
magnetic  white  dwarfs.   In  particular,  we  do  not  have  a  full
evolutionary picture of the progenitors  of magnetic white dwarfs. The
two main hypothesis are the following ones. Either magnetic fields are
inherited from  a weak magnetic field  of the progenitor star  --- the
so-called  fossil  field  hypothesis  --- or  are  originated  by  the
evolution in  a binary  system.  Both  hypothesis have  advantages and
drawbacks, and no definite consensus about this issue has been reached
so  far.   We discuss  them  in  detail in  Sect.~\ref{origin}  below.
Additionally,  there are  other  competing  scenarios which  challenge
those two  previously mentioned, which  are also examined in  the same
section.

Here we  briefly review the  current status and paradigms  of magnetic
fields in  white dwarfs, from  both the theoretical  and observational
points of view. However, we first  would like to draw the attention of
our reader  to the excellent  and recent review  of Ref.~\refcite{BF},
where  a  very  thorough  and  in depth  examination  of  our  current
understanding of the  research field was done.  Our  work is organized
as follows.  Sect.~\ref{observations} is devoted to summarize the most
relevant observational  characteristics of the population  of isolated
magnetic  white dwarfs.   Specifically,  in  Sect.~\ref{massd} we  pay
attention to the mass distribution  of magnetic white dwarfs, while in
Sect.~\ref{rotation}   we  discuss   their  rotational   periods.   In
Sect.~\ref{origin}  we critically  review  the  proposed scenarios  to
explain    the   presence    of    magnetic    fields.    Later,    in
Sect.~\ref{applications}  we  elaborate  on   some  of  the  practical
applications  of  the  field.    Finally,  in  Sect.~\ref{summary}  we
summarize  the main  results and  we propose  some interesting  future
research lines. Before going into details  we would like to state that
the  selection  of  papers  for  explicit  citation  may  be  somewhat
incomplete, for  several reasons. The first  one is that the  field is
rapidly evolving, the second one  is because of space limitations, and
finally the  last reason is that  this selection of references  is the
product of the  own special research trajectory of  the authors. While
we have tried  to be as complete  as possible, we are  well aware that
this has  not been possible.  Hence,  we apologize in advance  for any
unintentionally missed references.

\section{Observations of magnetic white dwarfs}
\label{observations}

Most magnetic white  dwarfs belong to the so-called  DA spectral type,
that is  they have hydrogen-rich  atmospheres. The reason for  this is
that  more than  80\% of  normal white  dwarfs also  belong to  the DA
spectral class.  Thus, the measurements of the magnetic field strength
in the majority  of the cases rely on  the spectroscopic determination
of the Zeeman  splitting of the Balmer series of  hydrogen.  It can be
easily proved that  for sufficiently low magnetic  field strengths the
splitting  of  these lines  depends  linearly  on the  magnetic  field
strength.   This allows  to place  upper  limits on  the existence  of
magnetic fields as  small as $10^5$~G. However, as  the field strength
increases  non-linear terms  become more  and more  important and  the
determination  of   the  field  strength  becomes   more  complicated.
Specifically, quadratic terms are important for field strengths of the
order  of $\sim  1$~MG, and  for larger  magnetic field  strengths the
situation becomes even  more complicated, as the  subcomponents of the
spectroscopic lines intermix in wavelenght.  Nevertheless, as of today
we have a handful of isolated  magnetic white dwarfs for which we have
reliable  determinations  of  magnetic  field strengths  as  large  as
800~MG\cite{BF}. Another technique frequently  used, because its value
in detecting  very strong magnetic  fields, consists of  measuring the
continuum   circular   polarization\cite{Kemp1970}.    However,   this
technique is  demanding observationally and  is only useful  for white
dwarfs with magnetic fields strengths exceeding $10^8$~G.

Observations  show  that the  population  of  isolated magnetic  white
dwarfs has two significant general  properties.  The first of these is
that apparently  there is  no clear  correlation between  the magnetic
field strength and the effective temperature. This would mean that the
field does not evolve appreciably along the white dwarf cooling track.
Nonetheless, this is still a controversial issue. In particular, it is
worth   mentioning   that  recent   observations\cite{kepler13}   have
demonstrated  that   the  mean   field  increases  at   the  effective
temperature  at  which  the   partially  degenerate  envelope  becomes
convective.   Whether  this  effect is  significant  deserves  further
scrutiny. The  fact that the field  does not evolve along  the cooling
sequence (if indeed this is the  case) can be well explained by simply
computing the ohmic  timescale.  This timescale is  defined as $t_{\rm
ohm}\sim 4\pi\sigma  L^2/c^2$, where $L$  is the typical  scale length
for the variation of the magnetic  field inside the star, and $\sigma$
is the electric conductivity. Adopting  $L\simeq R$ and typical values
for $\sigma$, it  can be shown\cite{Cumming} that  the ohmic timescale
is  indeed very  long  (of  the order  of  $10^{11}$~yr).  The  second
important general property is that  the topology of the magnetic field
can be  very complicated\cite{PJ95,Euchner1,Euchner2}  in most  of the
cases. However,  in practice and  for the  sake of simplicity  when no
more information is  available, it is customary to  simply assume that
the field geometry is dipolar.

There are as  well some magnetic white  dwarfs with hydrogen-deficient
atmospheres.   A few  of them  have helium-rich  atmospheres and  show
He{\sc I}  lines in their  spectra. For  these white dwarfs  the field
strength is  thus determined  using atomic helium  lines, a  much more
difficult task\cite{Jordan98}.  There is another group of white dwarfs
with   hydrogen-deficient   atmospheres    with   significant   carbon
abundances, the so-called DQ white  dwarfs. In some of these enigmatic
white dwarfs field strengths of the order of 100~MG have been measured
using spectropolarimetry\cite{Liebert1978}.  However,  this is not the
only   class   of   white    dwarfs   which   show   enhanced   carbon
abundances. Besides this  group of white dwarfs  which are essentially
cool, there  is a  recently discovered\cite{Dufour2007}  population of
hot  DQs, which  have carbon-dominated  atmospheres. It  is intriguing
that about half of them are magnetic white dwarfs. Finally, there is a
distinct group  of magnetic white  dwarfs, known as magnetic  DZ white
dwarfs,  with metals  in their  atmospheres\cite{Hollands,Kepler2015b,
keplernew}.

\subsection{Mass distribution of magnetic white dwarfs}
\label{massd}

The determination of masses of magnetic  white dwarfs is a tough task,
because of the inherent difficulties  in modeling the line profiles in
the presence of  a magnetic field. In particular,  we lack theoretical
models  allowing to  model  accurately pressure  broadening for  large
magnetic  fields.   According  to  this consideration,  we  only  have
reliable  mass determinations  for a  reduced subset  of all  magnetic
white dwarfs\cite{Kulebi10}, whereas for  most high field white dwarfs
the   mass   determination   is  somewhat   uncertain.    However,   a
characteristic  trend emerges  from observations.   In particular,  it
turns out that the average  mass of isolated high-field magnetic white
dwarfs --- namely, those with magnetic  fields larger than 1~MG --- is
substantially  larger   than  that  of  single   field  white  dwarfs.
Specifically, the population of high-field magnetic white dwarfs has a
mean mass of $0.784\pm 0.047\, M_{\odot}$,\cite{Silvestri2007} whereas
the  average mass  of single  non-magnetic white  dwarfs is  $0.643\pm
0.136\,  M_{\odot}$,\cite{Tremblay2013}  clearly  pointing  towards  a
different evolutionary channel for these white dwarfs.

\subsection{Rotational periods}
\label{rotation}

The vast majority of magnetic white dwarfs rotate slowly, as it occurs
for non-magnetic  white dwarfs\cite{Brassard2008}.   Specifically, the
rotation periods  of isolated magnetic  white dwarfs encompass  a wide
interval\cite{Jordan2002,Brinkworth},  with  a  lower limit  of  $\sim
700$~s, whereas for  some magnetic white dwarfs  the measured rotation
periods are actually  much longer, on the order of  about 100~yr. As a
matter of fact,  there is weak evidence for a  bimodal distribution of
rotation periods, with a handful of magnetic white dwarfs with periods
clustered around hours,  and a second (more numerous)  subset of stars
with  periods  much longer  than  this  value, typically  hundreds  of
years. We note that  this is a crucial issue, since  it would allow us
to  discern  the  progenitors  of  magnetic  white  dwarfs  (see  next
section).  However, until now  the intrinsic difficulties of measuring
accurate periods  using photometry  and polarimetric  variability have
hampered the efforts to provide a definite answer to this problem.

\section{The origin of the magnetic field}
\label{origin}

The search for  the progenitors of magnetic white dwarfs  is an active
field of  research, and unfortunately  no consensus on this  issue has
been  reached yet.   Generally  speaking,  the evolutionary  scenarios
giving rise to the known population of magnetic white dwarfs should be
able to explain three well established observational features of their
ensemble properties.  The first one  is that high-field magnetic white
dwarfs are usually more massive than their non-magnetized counterparts
--- see Sect.~\ref{massd}.  The second important observational fact is
that  most   magnetic  white   dwarfs  are   slow  rotators   ---  see
Sect.~\ref{rotation}.    Finally,   there   is   another   interesting
observational fact  that deserves  close attention.   For non-magnetic
white dwarfs there exists a well known population of binaries in which
one of  the members  of the  pair is a  main-sequence star,  while the
other  one   is  a  white  dwarf\cite{ARM13}.    Realistic  population
synthesis models are able to reproduce the most relevant properties of
well  characterized  samples\cite{Camacho1,Camacho2}.    Thus,  it  is
commonly assumed that  the scenarios that produce  this population are
relatively well  understood, although  much work  still remains  to be
done.   However,  magnetic  white   dwarfs  are  predominantly  single
stars\cite{Liebert2015}. Even  more, it is found  that, surprinsingly,
the  white dwarf  companion in  cataclysmic variables  is magnetic  in
about 25\% of  the systems.  All this strongly  suggests that binarity
plays a key role in explaining the origin of at least some fraction of
the presently observed population of magnetic white dwarfs.  There are
two competing scenarios which may  eventually explain the formation of
magnetic white dwarfs.  These are  the fossil field hypothesis and the
binary scenario. In the following we examine them separately.

\subsection{The fossil field hypothesis}
\label{fossil}

We  start  describing   the  fossil  field  hypothesis\cite{Angel1981,
WF2005}.  Within this evolutionary channel the magnetic field of white
dwarfs  is  simply  the  consequence  of the  evolution  of  a  single
progenitor  along  all  the   standard  stellar  evolutionary  phases.
Specifically, within this scenario  magnetic white dwarfs descend from
rotating Ap  and Bp stars, which  are the only class  of main-sequence
stars known  to have substantial  magnetic fields, between  $10^3$ and
$10^5$~G. If  the effects of  mass loss  are neglected and  we further
assume that  magnetic flux is  conserved it is  easy to show  that the
field will be amplified by a factor of $\sim 10^4$ when the progenitor
becomes a  white dwarf.  Even  if the  assumptions are relaxed,  and a
significant  amount of  magnetic flux  is  carried away  by mass  loss
during advanced evolutionary stages before a white dwarf is formed, it
is expected that the magnetic field of the resulting white dwarf would
be  comparable to  those  typically found  in  magnetic white  dwarfs.
However, this scenario  faces a serious drawback.  Specifically, it is
not  able to  explain  why  there are  not  magnetic  white dwarfs  in
post-common  envelope  binaries  with  a  main-sequence  companion  of
spectral type K or M.

\subsection{The binary hypothesis}
\label{binary}

Within  the  second scenario,  the  so-called  binary hypothesis,  the
magnetic  field arises  from the  interaction of  the future  magnetic
white  dwarf with  a companion  during its  previous evolution.   This
evolutionary channel  has been the  subject of much  recent attention,
and there are several variants of  the scenario.  For instance, it has
been suggested\cite{Tout2008, Nordhaus2011} that strong magnetic fiels
are produced during a common envelope episode in a close binary system
in which one of the components of the pair is degenerate.  During this
phase, spiral-in of the secondary induces differential rotation in the
extended  convective  envelope, resulting  in  a  stellar dynamo  that
produces   the   magnetic   field.    However,  it   has   also   been
shown\cite{Potter2010} that  the magnetic  field produced in  this way
does not  penetrate into the white  dwarf, and it decays  rapidly when
the  common   envelope  is  ejected.    It  has  been   also  recently
demonstrated\cite{mio}   that   the   hot,   differentially   rotating
convective  corona  resulting  from   the  merger  of  two  degenerate
cores\cite{Pablo} produces strong magnetic fields that are confined to
the outer  layers of the resulting  remnant and do not  decay for very
long   timescales.   Indeed,   detailed  three-dimensional   numerical
simulations\cite{Ji} using state-of-the-art  computer codes have shown
that the  a very small magnetic  field is amplified during  the merger
episode   and   that  the   remnant   of   the  merger   is   strongly
magnetized. Hence, this evolutionary channel would explain some of the
gross  properties  of  the  population of  high-field  magnetic  white
dwarfs. It  might be argued  that this  scenario might be  in conflict
with the observational  fact that most magnetic white  dwarfs are slow
rotators. However,  it has  been also  recently shown\cite{Kulebi2013}
that  coupling  between  the   magnetosphere  and  the  debris  region
resulting from the disruption of  the secondary star during the merger
episode can brake the magnetized  white dwarf and bring the rotational
periods to values comparable  to those observationally found. Finally,
we mention that there  is additional observational evidence supporting
the   binary   hypothesis.     In   particular,   recent   large-scale
searches\cite{Asensio2014}  for magnetic  central  stars of  planetary
nebula and hot subdwarfs have  confirmed previous suggestions that the
these  stars are  basically non-magnetic\cite{Mathys2012,  Landstreet,
Jordan12}.

\subsection{Population synthesis studies}
\label{population}

\begin{figure}[t]
\begin{center}
\includegraphics[width=1.0\textwidth]{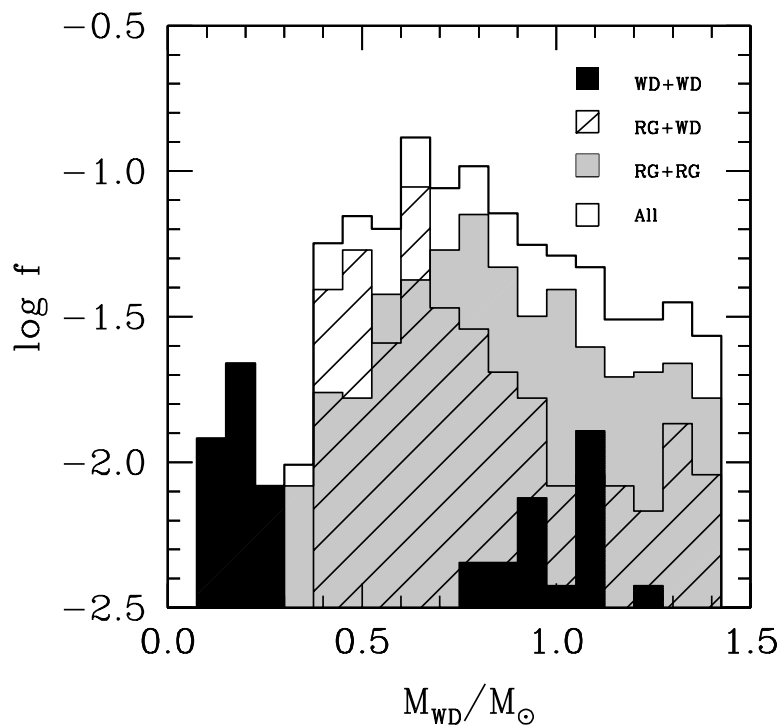}
\caption{Mass distribution of the  remnants of several merger channels
  in the  Solar neighborhood, from Ref.~29.   The different histograms
  show  the   frequency  of  the  merger   channels  considered  here.
  Specifically, the black  histogram shows the masses  of the remnants
  of the mergers of double  white dwarf binaries, the dashed histogram
  that of the mergers of a binary system composed of a red giant and a
  white dwarf,  the shaded histogram  that of  the mergers of  two red
  giants, while  the total  mass distribution is  shown using  a solid
  line.}
\label{mergers}
\end{center}
\end{figure}

Population  synthesis studies  are crucial  to discern  the origin  of
magnetic white  dwarfs.  It  is important  to do  this because  in the
first  case the  number of  mergers in  the Solar  neighborhood is  an
important  piece  of  evidence  in determining  if  this  evolutionary
channel  might provide  enough progenitors  to explain  the number  of
high-field magnetic white  dwarfs in a volume  limited sample, whereas
for the second case  the number Ap and Bp stars,  due to its intrinsic
scarcity,  may not  be sufficient  to explain  the fraction  of single
magnetic  white  dwarfs.   According   to  these  considerations,  the
predictions  of such  studies for  both  the binary  scenario and  the
fossil field  evolutionary channel have been  compared to observations
in recent years.  Moreover, dedicated  surveys have provided us with a
number of binary systems which  potentially will merge within a Hubble
time, and  this can  be directly  compared to  the predictions  of the
theoretical models,  allowing in this way  to test our models  for the
binary scenario.

We start this section by examining  the statistics of the fossil field
evolutionary channel.  As a  matter of fact,  it was  early recognized
that  the  number of  strongly  magnetic  Ap  and  Bp stars  could  be
insufficient to explain  the observed incidence of  magnetism in white
dwarfs. Recent studies\cite{Kawka} have  argued that these stars could
not  be the  only progenitors  of magnetic  white dwarfs,  because the
birth  rate of  such stars  is  not enough  to explain  the number  of
observed   magnetic   white   dwarfs.    Nevertheless,   more   recent
studies\cite{WF2005} have  concluded that  this problem can  be easily
overcome by taking  into account that about 40\% of  late type B stars
have  undetectable   magnetic  fields.   This  fraction   of  magnetic
main-sequence  stars   would  be   enough  to  reconcile   theory  and
observations of magnetic  Ap and Bp stars with  distances smaller than
100~pc.\cite{Power2008} Moreover, it is observationally found that the
incidence of magntism  in A and B stars increases  with mass. All this
precludes from discarding the fossil field evolutionary channel.  Even
more, it is  quite plausible that at least some  magnetic white dwarfs
have this kind of progenitors.

We now  turn our  attention to  the binary  scenario. The  most recent
studies  of  this kind\cite{Briggs,mio}  agree  in  predicting that  a
sizable fraction  of magnetic  white dwarfs can  be explained  by this
scenario,   provided   that   several  types   of   coalescences   are
considered. To  better illustrate  this, Fig.~\ref{mergers}  shows the
frequency  distribution  of remnant  masses  of  the different  merger
channels  for  a   sample  of  $10^3$  mergers.    In  this  frequency
distribution   all  the   remnants   with  masses   larger  than   the
Chandrasekhar limiting  mass have been  removed.  As can be  seen, the
total mass  distribution (open  histogram) presents  a first  peak for
masses smaller  than $\sim 0.4\, M_{\odot}$,  corresponding to mergers
in which a helium white dwarf  is produced, then sharply increases for
increasing remnant masses and afterwards smoothly decreases for masses
larger than $\sim 0.6\,  M_{\odot}$. When the theoretical distribution
is sampled  for $\sim  14$ objects  --- the  total number  of magnetic
white dwarfs within 20~pc, see below --- fairly flat distributions are
obtained  for   masses  ranging  from  $0.8\,   M_{\odot}$  to  $1.4\,
M_{\odot}$.

We now discuss the statistics of the local sample. Within 20~pc of the
Sun there are 122 white  dwarfs\cite{Holberg2008}, and several of them
are  magnetic\cite{Kawka}.  This  sample is  80\% complete,  but still
suffers from poor statistics.  However, it is useful because for it we
have a  reliable determination of  the true incidence of  magnetism in
white  dwarfs.  Mass  determinations are  available for  121 of  these
white dwarfs, and there are 14 magnetic white dwarfs. Of these, 8 have
magnetic fields  larger than $10^7$~G,  and 3 have masses  larger than
$0.8\, M_{\odot}$ --- a value which  is $\sim 2.5\sigma$ away from the
average mass of field white dwarfs.  The selection of this mass cut is
somewhat  arbitrary  but, given  the  strong  bias introduced  by  the
initial  mass function,  it  is  expected that  the  vast majority  of
high-field magnetic  white dwarfs more massive  than $0.8\, M_{\odot}$
would be the result of stellar mergers.

The  population synthesis  calculations  predict that  $\sim 4$  white
dwarfs are  the result of  double degenerate mergers, and  have masses
larger than  $0.8\, M_{\odot}$,  in good agreement  with observations.
This has to be compared with the fraction of white dwarfs more massive
than $\sim  0.8\, M_{\odot}$ resulting from  single stellar evolution,
which is  $\sim 10\%$.  Consequently,  the expected number  of massive
white dwarfs  in the local sample  should be $\sim 12$.   Instead, the
local sample  contains 20, pointing  towards a considerable  excess of
massive white dwarfs, which could be the progeny of mergers.  The rest
of the  population of magnetic  white dwarfs  ($\sim 5$) could  be the
result of the evolution  of single stars\cite{AznarCuadrado, Blackman}
--- see above.

Finally, we mention that the  number of coalescing binaries previously
discussed compares well with the results obtained using very different
population  synthesis  codes. In  summary,  we  are confident  that  a
substantial fraction of high-field magnetic white dwarfs should be the
result of stellar mergers.

\subsection{Assessing the birth rates}
\label{rates}

The rate  of double degenerate  mergers has  been the subject  of much
attention recently, because of its  implications on different areas of
high  energy  astrophysics.   Among  them we  mention  explicitly  the
following ones.   The coalescence of  two white  dwarfs is one  of the
possible    scenarios   to    account    for    Type   Ia    supernova
outbursts\cite{Webbink,IT84}. It is thought as well that the merger of
two   degenerate    cores   could    lead   to   the    formation   of
magnetars\cite{King2001}.   Also,  the  hot and  massive  white  dwarf
members of the Galactic halo could be the result of the coalescence of
a   double   white-dwarf   binary   system\cite{Schmidt,   Segretain}.
Additionally,   hydrogen-deficient  carbon   and  R   Corona  Borealis
stars\cite{Izzard,   Clayton,  Longland}   are  thought   to  be   the
consequence of the merging of  two white dwarfs.  Also, the relatively
high photospheric metal abundances  of some hydrogen-rich white dwarfs
with circumstellar  disks around them  could also be explained  by the
merger    of   a    carbon-oxygen   and    a   helium    white   dwarf
\cite{GD362}. However, we note that  not all massive white dwarfs with
large metal abundances show significant infrared excesses, and thus it
is unlikely that they harbor disks around them\cite{Hansen}.  AM Canum
Venaticorum systems  are as well  thought to  be the consequence  of a
merger, as also are single subdwarf B/O stars\cite{Bildsten}. Last but
not  least,  the  phase  previous  to  the  coalescence  of  a  double
white-dwarf close binary system has been shown to be a powerful source
of  gravitational  waves  that   would  be  eventually  detectable  by
LISA\cite{GWR}. In this  section we review the status of  the field in
the context  of the scenarios  leading to the formation  of magnetized
white dwarfs.

The two most significant efforts to find close binary systems in which
both components of  the pair are degenerate are the  ESO Supernovae Ia
Progenitor Survey (SPY)\cite{Napi1, Napi2}  and the Extremely Low Mass
(ELM) white dwarf  survey\cite{Kilic2010, Brown2010, Brown2015}.  Both
are  dedicated  surveys,  and adopt  different  observing  strategies.
Specifically, the  SPY survey is  a magnitude limited survey  aimed at
searching for double-degenerates,  whereas the primary aim  of the ELM
survey is  to search  for binary systems  containing a  low-mass white
dwarf.  Both  surveys have  provided us with  an invaluable  wealth of
observational   data,   consisting   of    several   dozens   of   new
double-degenerate systems, which  allows us to compare  the results of
the  population synthesis  models described  before with  the observed
distributions.   To   the  findings  of  these   surveys  the  several
double-degenerate systems  found serendipitously  in the SDSS  must be
added.

In synthesis, the main result  of these observational efforts is that,
as of today,  none of the surveys  has been able to  find a progenitor
system for Type Ia supernovae.  That is, none of the surveys has found
yet  a double-degenerate  system with  a  total mass  larger than  the
Chandrasekhar  limiting mass  that will  merge in  less than  a Hubble
time.   However, there  are other  interesting results  that are  more
suitable  for our  interests. We  focus primarily  on the  ELM survey,
because  it is  the most  recent  one.  The  observed distribution  of
periods  peaks   at  around   half-a-day,  and  follows   a  lognormal
distribution\cite{Andrews}.   Also,  the   mass  distribution  of  the
low-mass companion peaks at $\sim  0.2\, M_{\odot}$ with a very narrow
dispersion,  whereas the  mass distribution  of the  massive companion
follows a Gaussian law, which peaks at $\sim 0.75\, M_{\odot}$, with a
relatively  large dispersion  of about  $\sim 0.25\,  M_{\odot}$.  The
estimated  birth  rate  of  these  systems,  once  corrected  for  the
observational  biases  and  selection   effects,  is  $\sim  4.0\times
10^{-3}$~yr$^{-1}$.  However, the birth rate  of systems that will end
up  their evolution  as R~Coronae  Borealis stars  is $\sim  3.0\times
10^{-3}$~yr$^{-1}$,\cite{Brown2014,  Zhang2014} those  giving rise  to
AM~Canum      Venaticorum     systems      is     $\sim      1.0\times
10^{-4}$~yr$^{-1}$,\cite{Carter} and that of systems that will produce
underluminous        supernovae        is       $\sim        1.0\times
10^{-4}$~yr$^{-1}$.\cite{Foley2009}  Thus, the  birth rate  of systems
that would  eventually produce  magnetic white dwarfs  is considerably
smaller,  $\sim  8.0\times  10^{-4}$~yr$^{-1}$.  These  estimates  are
consistent with the theoretical expectations described in the previous
section,  and  with   the  merger  rates  derived   by  analyzing  the
SDSS\cite{Badenes}.

A tantalizing possibility  is that some R~Coronae  Borealis stars turn
into magnetic  white dwarfs.  Whether  this is possible remains  to be
assessed. However, if  true, the contribution of mergers  to the birth
rate of magnetic  white dwarfs may be higher  than previously thought,
and   certainly  larger   than  the   current  estimates,   $8.0\times
10^{-4}$~yr$^{-1}$.   Given the  $4.0\times 10^{-3}$~yr$^{-1}$  merger
rate from the ELM Survey, it  can argued that there are enough mergers
in the Solar neighborhood to explain magnetic white dwarfs.

\subsection{A new scenario}
\label{convection}

Finally, we mention that a  new evolutionary scenario is progressively
emerging\cite{nos}.  The basic assumption  within this scenario hinges
on the observational  fact that the fraction of  magnetic white dwarfs
seems  to  increase  for  decreasing  luminosities\cite{Gianmichelle}.
Thus,  it  could  be  well   possible  that  the  magnetic  fields  of
low-magnetized  white  dwarfs  could  be  originated  by  an  internal
physical process.   Specifically, the number of  single magnetic white
dwarfs increases  abruptly for  luminosities $\log(L/L_\odot)\la-3.5$.
Interestingly,  for an  otherwise typical  white dwarf  of mass  $\sim
0.6\, M_\odot$  this luminosity corresponds  to a core  temperature of
$\sim 10^6$~K, which is the  temperature at which crystallization sets
in\cite{Review,Renedo}.   This strongly  suggests that  the convective
mantle\cite{isern1,isern2}  that  results   from  carbon-oxygen  phase
separation\cite{GB1,GB2} upon crystallization  would produce a stellar
dynamo  resembling closely  that occurring  in the  interior of  Solar
system planets\cite{Christensen}.  Actually, it  can be shown that the
energy involved in the Rayleigh-Taylor  unstable region is the same as
what  is needed  to explain  low-field magnetic  white dwarfs,  namely
those with magnetic field strenghts smaller than 0.1~MG.  Nonetheless,
this is still a preliminary model that should be further developed and
their predictions should be compared with observations.

\section{Applications}
\label{applications}

In   addition  to   their   obvious  and   mumerous  applications   to
astrophysical phenomena occurring in cataclysmic variables, the theory
of magnetic white  dwarfs also has many  interesting applications that
deserve to be mentioned. In the following we detail some of them. 

\subsection{Anomalous X-ray pulsars}

One of  the possible applications  of these  types of studies  is that
high-field magnetic  white dwarfs could  explain the properties  of at
least a  fraction of anomalous  X-ray pulsars.  This class  of pulsars
shares some similarities with short gamma-ray repeaters, which radiate
short  ($\approx  100$~ms), repeating  bursts  of  soft $\gamma$-  and
X-rays at  irregular intervals. In particular,  their rotation periods
cluster  between 2  and 12~s,  have  large magnetic  fields, and  have
quiescent     X-ray      luminosities     of     the      order     of
$10^{35}$~erg~s$^{-1}$. The traditional explanation for this enigmatic
class of  pulsars is  that they are  magnetars\cite{Duncan1, Duncan2}.
Nevertheless,  there  are  alternative   models  that  challenge  this
picture.   In  particular,  following  early  suggestions\cite{Morini,
  Paczynski} it  has been recently proposed\cite{Malheiro}  that these
sources can be better explained assuming  that the object is a rapidly
rotating      highly     magnetized      white     dwarf.       Recent
calculations\cite{Rueda} have  shown that  this model can  explain the
properties of the anomalous X-ray  pulsar 4U~0142+61, thus making this
a very suggestive formation scenario for these objects.

\subsection{Millisecond pulsars}

Millisecond pulsars are  a distinct subset of the  known population of
pulsars. They have  magnetic fields with stregths  ranging from $10^8$
to $10^9$~G, somewhat  smaller thant the rest of  radio pulsars, which
have magnetic fields up to  $10^{13}$~G. Moreover, they are frequently
found  in  binary  systems.   Actually,   $\sim  75\%$  of  them  have
companions\cite{Hurley}.   

It is  generally accepted that  millisecond pulsars are  neutron stars
that were  originated in a  core-collapse supernova event in  a binary
system. Within this  evolutionary scenario the massive  star that will
eventually yield the  newborn neutron star accretes  material from the
companion,  and   the  system   is  detected   as  a   low-mass  X-ray
binary\cite{Bhattacharya}.   In  this  case   the  magnetic  field  is
originated by the standard recycling hypothesis. That is, the field is
constrained  deep in  the superconducting  core of  the neutron  star.
However,  the large  prevalence  of low-field  millisecond pulsars  in
binary  system has  brought  into question  the standard  evolutionary
scenario. One of the possibilities is that these pulsars are formed by
accretion-induced collapse of an oxygen-neon white dwarf\cite{Michel}.
In this  evolutionary route an  accreting massive white dwarf,  with a
core made of oxygen and neon,  reaches the threshold density to enable
electron  captures on  $^{24}$Mg  and $^{24}$Na  first,  and later  on
$^{20}$Ne  and $^{20}$F,  to finally  ignite Ne  and O  explosively at
central densities higher than  $\sim 2\times 10^{10}$~g~cm$^{−3}$.  At
these very high central densities,  fast electron captures occuring on
the nuclear  statistical equilibrium material would  rapidly drive the
Chandrasekhar mass below  the actual mass of the  degenerate core and,
consequently,  gravitational  collapse  would ensue,  leading  to  the
formation of a neutron star\cite{Guti1, Guti2}.  Within this formation
route  white dwarfs  with initially  small magnetic  fields (of  about
$10^4$~G)  can  explain naturally  the  observed  properties of  these
pulsars, by  simply assuming that  the magnetic field is  amplified by
flux  conservation. The  observational counterparts  in this  scenario
would also be low-mass X-ray binaries.

Recent  population synthesis  studies\cite{Hurley} have  examined both
possibilities  and  have  concluded  that  the  birthrates  of  binary
millisecond  pulsars  formed  through accretion-induced  collapse  are
comparable to and can exceed those  for core collapse, but these types
of studies are not yet conclusive,  so clearly more efforts need to be
pursued.

\subsection{Magnetic double degenerates}

Magnetic double  degenerates are rare  systems, especially if  the two
degenerate stars form a wide  pair. However, these binary systems have
the  advantage of  allowing us  to study  the origin  of the  magnetic
field.  In these systems the components of the binary are sufficiently
separated to have evolved independently, so  the age of the system and
the   distance   can   be    evaluated   studying   the   non-magnetic
companion\cite{Girven}.  However,  very few  systems of this  type are
known.   Among  them  we  mention the  following  ones.   RE~J0317-853
\cite{Barstow95}, was discovered  by ROSAT and is  relatively close to
us\cite{Kulebi10}, thus allowing for  accurate measurements.  The pair
is composed  by a massive white  dwarf of mass $\sim  0.85\ M_{\odot}$
and an ultramassive  white dwarf of unknown mass which  has a magnetic
field of $\sim 450$~MG.\cite{Burleigh99}.  Another example is the pair
formed by PG~1258+593 and SDSS~J130033.48+590407.0.  In this case both
white   dwarfs  have   nearly  equal   normal  masses,   $\sim  0.54\,
M_\odot$. The magnetic component has a  field strength of 6~MG, and is
the  cool component  of the  system.  To  them we  add two  recent new
discoveries: SDSS  J092646.88+132134.5 + J092647.00+132138.4  and SDSS
J150746.48+521002.1 + J150746.80+520958.0.  The  white dwarfs in these
systems are more  massive than usual in field white  dwarfs. All these
binary systems  are common  proper motion  pairs.  However,  there are
also systems for  which the components are not well  resolved. This is
the   case   of  LB~11146   which   we   know   is  a   close   binary
system\cite{Liebert93,  Nelan2007},   a  characteristic   shared  with
similar    systems,    like     RE~J1439+75,    \cite{Vennes99}    and
G62--46.\cite{Bergeron93} With these very  few systems it is difficult
to reach definite conclusions, but this  is a promising line of future
research.

\subsection{Magnetic white dwarfs and Type Ia supernovae}

Type Ia supernovae  are one of the most energetic  explosive events in
the  cosmos.  Since  there  is a  relationship  linking its  intrinsic
brightness  and the  shape  of  their light  curves  and  they can  be
detected at  very large distances  they can be used  as standardizable
cosmological candles.  This has opened a new era in cosmology, and has
enabled  us  to  discover  the acceleration  of  the  universe  (\cite{Riess_1998,  Perlmutter_1999}),  
and  to  determine  the  cosmological
parameters.

Despite  their importance,  we still  do not  know the  nature of  the
progenitors  of  Type Ia  supernovae,  which  remains a  long-standing
mistery. We do know that the outburst is powered by the explosion of a
carbon-oxygen white dwarf  in a binary system, but we  do not know the
precise  mechanism  that destabilizes  the  white  dwarf, and  several
hypothesis have been put  forward.  In the so-called single-degenerate
channel, accretion  from a  non-degenerate companion onto  the primary
companion leads  to the formation and  explosion of Chandrasekhar-mass
white dwarfs.  However, recent  observational evidence suggests that a
diversity of progenitors exists, including a significant population of
sub-Chandrasekhar         and         super-Chandrasekhar         mass
systems\cite{ruiteretal11,  scalzoetal14}.    Therefore,  alternatives
have been proposed.  The most widely accepted competing model consists
of  the merger  of a  binary white  dwarf system\cite{Webbink,  IT84}.
This is  known as the  double-degenerate channel.  However,  there are
other  alternative  scenarios.    These  include  the  core-degenerate
channel\cite{sparksstecher74,        livioriess03,       kashisoker11,
ilkovsoker13,  aznarsiguanetal15},  and  the  white-dwarf  collisional
scenario\cite{raskinetal09, rosswogetal09,  thompson11, kushniretal13,
aznarsiguanetal13}.   Here,  for  obvious  reasons, we  focus  on  the
double-degenerate channel.

The double-degenerate channel offers natural explanations to a variety
of  observational facts,  including the  absence of  H$\alpha$ in  the
nebular     phase\cite{leonard07},     and      the     delay     time
distribution\cite{maozbadenes10, maozetal10}.   However, this scenario
also has several  major shortcommings that need to  be addressed.  The
most recent  theoretical works  have paid  attention primarily  to the
violent    merger    mechanism   (\cite{pakmoretal10,    pakmoretal11,
pakmoretal12, raskinetal12,  danetal12}).  This  mechanism is  based on
the behavior  found in  extensive numerical  simulations of  the final
phases of the  coalescence. During these phases the  secondary star is
tidally disrupted  and is rapidly accreted  onto the primary in  a few
dynamical timescales.   In contrast,  the primary star  remains almost
intact.   However, not  all the  mass  of the  disrupted secondary  is
accreted onto  the primary.  In  fact, all simulations predict  that a
hot, virialized accretion disk surrounding the primary, with a mass of
about  half of  the  mass  of the  secondary,  is formed  \cite{Pablo,
schwabetal12, Zhu_2013, danetal14}, while the remaining mass is indeed
accreted and forms a hot, convective corona\cite{mio}.  This region is
prone to  magneto-rotational instability.  The early  suggestions that
this mechanism  could give rise  to powerful magnetic fields  has been
recently confirmed  using full  three-dimensional magneto-hydrodynamic
calculations\cite{Ji,rahul,pakmor15},   but  unfortunately   the  only
simulations done so far do  not encompass massive enough white dwarfs,
a requisite to produce a powerful detonation\cite{Seitenzahl_2009}.

Magnetic fields  most likely  play a crucial  role in  explaining some
properties of  Type Ia  supernovae. However, despite  this potentially
important interest,  very few studies  have addressed this  issue, and
much work still remains to  be done. For instance, the characteristics
of some overluminous supernovae, with nickel masses larger than $1.0\,
M_{\odot}$ like SN  2003fg, SN 2006gz, SN 2007if and  SN 2009dc, might
be   explained   if   a   sufficiently   large   magnetic   field   is
present\cite{Das12,Das13}.        However,        modeling       these
super-Chandrasekhar explosions  requires taking into account  not only
the  effects  of   the  magnetic  pressure,  but   also  dealing  with
general-relativistic corrections. A full treatment of these issues has
only recently been done\cite{Das14,Das15},  and although this research
line is promising more theoretical  calculations are needed to confirm
the  results  obtained  so  far.  Finally,  we  mention  that  another
possibility   has   arised    recently\cite{Becerra}.    Namely,   the
post-merger evolution  of the  coalescences with  a total  mass larger
than the  Chandraskhar limit  could be dominated  by the  magnetic and
accretion torques. Thus,  a delayed explosion of  the central spinning
white  dwarf would  be possible.  This detonation  would be  caused by
magnetic braking.

\section{Summary and outlook}
\label{summary}

In this paper  we have reviewed our current  understanding of magnetic
white dwarfs.   This class of  objects, is interesting not  only ``per
se'',  but also  for its  many and  interesting applications  in other
areas  of contemporary  astrophysics.   Some of  them  have also  been
reviewed here. The current  observational sample comprises about $\sim
250$ objects for which we have reliable determinations of the magnetic
field  strength, and  for  several  of them  we  also have  relatively
accurate mass  determinations. However,  it is worth  emphasizing that
probably there are more magnetic white dwarfs with low field strengths
for which the current limitations of the observational techniques have
not allowed  us to determine the  strength of the magnetic  field, and
thus  there is  quite likely  a  hidden population  of magnetic  white
dwarfs  with very  low  magnetic field  strengths.  Nevertheless,  the
existing wealth  of observational  data ---  primarly provided  by the
recent, advanced large-scale  surveys, like the SDSSS  --- is nowadays
being  analyzed. This  includes not  only studying  the properties  of
individual objects, but  also deriving the ensemble  properties of the
population of magnetic white dwarfs. This last analysis has allowed us
to unveil two sub-groups of stars. First, there is a group of magnetic
white dwarfs  with moderately  low magnetic  fields which  have masses
close  to  the average  of  their  non-magnetic analogs.   The  second
sub-group consists of a distinct set of massive white dwarfs with very
high  magnetic fields,  typically  of the  order  of $10^9$~G.   These
observational advances have yielded some  insight on the origin of the
magnetic  fields, but  still there  is much  work to  be done  in this
respect, and clearly  theoretical models need to be  improved to match
observations. In  particular, we stress  that there are  two competing
theoretical scenarios for the formation  of magnetic white dwarfs, and
there  is not  yet enough  concluding evidence  favoring one  of them.
However,  it is  also  true that  the field  has  advanced in  several
distinct ways since the discovery of the first magnetic white dwarf.

Certainly, the next decade will see  a dramatic increase in the number
of  known  magnetic  white  dwarfs.  Future  releases  of  large-scale
surveys,  like astrometric  satellite Gaia\cite{Gaia}  or that  of the
Large  Synoptic  Survey  Telescope (LSST)  project\cite{Saha09},  will
definitely  allow us  to find  many variable  white dwarfs,  including
magnetic  white dwarfs  with spots.   However,  in the  case of  Gaia,
ground-based follow-up  spectroscopy of these objects  will be crucial
to getting  the most out  of the  Gaia observations. With  a geometric
parallax accuracy  of 1 milli-arcsecond  and very deep  exposures, the
LSST  parallax survey  will  match the  faint-end  precision of  Gaia,
providing a nearly complete catalog (including accurate parallaxes) of
white  dwarfs up  to  $M_{\rm v}  =  15$ in  selected  regions of  the
southern sky. Moreover, it is  foreseen that a significant fraction of
them will be magnetic.  By  analyzing these samples with model spectra
we expect to have a much clearer picture of the population of magnetic
white dwarfs.   More than anything  else, these enhanced  samples will
undoubtely constitute  important tools  for unraveling the  origin and
evolution of magnetic fields in stars.

\section*{Acknowledgments}

E.G-B  gratefully acknowledges  the  financial support  of the  MINECO
grant  AYA2014--59084--P,   and  of   the  AGAUR.    M.K.   gratefully
acknowledges the support of the  NSF and NASA under grants AST-1312678
and  NNX14AF65G,  respectively.  Partial financial  support  for  this
research comes from CNPq (Brazil).

\end{document}